\documentclass[12pt,preprint]{aastex}

\begin{document}

\slugcomment{Accepted to ApJ 4/7/06.}

\title{Laboratory Testing of a Lyot Coronagraph Equipped with an Eighth-Order Notch Filter Image Mask}

\author{Justin R. Crepp, Jian Ge, Andrew D. Vanden Heuvel, Shane P. Miller}
\affil{Astronomy Department, University of Florida \\
211 Bryant Space Science Center, P.O. Box 112055 \\
Gainesville, FL 32611-2055}
\email{jcrepp@astro.ufl.edu, jge@astro.ufl.edu, avh@astro.ufl.edu, smiller@astro.ufl.edu}

\author{Marc J. Kuchner}
\affil{Exoplanets \& Stellar Astrophysics Laboratory \\
NASA - Goddard Space Flight Center \\
Greenbelt, MD 20771}
\email{marc.kuchner@nasa.gov}

\begin{abstract}
We have built a series of notch filter image masks that make the
Lyot coronagraph less susceptible to low-spatial-frequency optical
aberrations. In this paper, we present experimental results of their
performance in the lab using monochromatic light. Our tests show
that these ``eighth-order'' masks are resistant to tilt and focus
alignment errors, and can generate contrast levels of $2 \times
10^{-6}$ at 3 $\lambda / D$ and $6 \times 10^{-7}$ at 10 $\lambda /
D$ without the use of corrective optics such as deformable mirrors.
This work supports recent theoretical studies suggesting that
eighth-order masks can provide the Terrestrial Planet Finder Coronagraph
with a large search area, high off-axis throughput, and a practical
requisite pointing accuracy.
\end{abstract}


\section{INTRODUCTION}
\label{sec:intro}

A coronagraph is an instrument that controls the diffracted light
from a bright astrophysical object in order to image faint off-axis
features in its immediate vicinity. Future space-based missions,
such as the Terrestrial Planet Finder (TPF) mission, will use a
stellar coronagraph to search for Earth-like planets orbiting in the
habitable zone of nearby stars (Ford et al. 2004). Demonstrating
this technology requires suppressing broadband visible starlight by
more than a factor of $\sim 10^{10}$ only a few diffraction widths from
the telescope optical axis. In this paper, we report on progress
towards achieving this goal in the lab with a Lyot coronagraph that
is equipped with an ``eighth-order'' notch filter image mask.

The Lyot coronagraph consists of an entrance aperture, an image
mask, a field stop, and a detector \cite{lyot}. Light entering the
coronagraph is focused onto the mask in the first image plane. If the 
telescope is pointed accurately, most of the starlight
is occulted by the mask, while off-axis sources suffer little
attenuation. The light is then diffracted to a pupil plane, the Lyot
plane, where a stop is used to block the remaining starlight (Fig.
1).

\begin{figure}[!ht]
\centerline{
\includegraphics[height=1.4in]{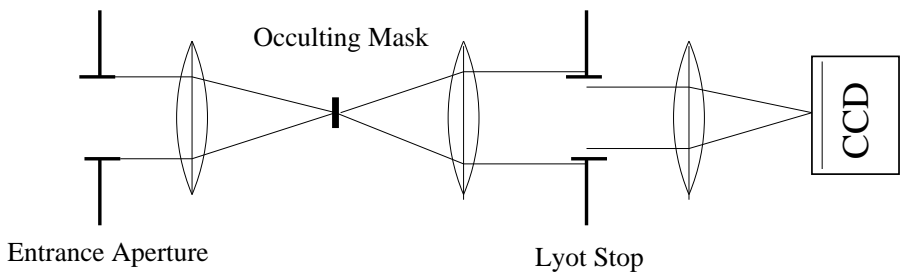}
} 
\caption{Optical layout of a transmissive Lyot coronagraph without
wavefront correction. See Sivaramakrishnan et al. 2001 for a
first-order theory description of how light interacts with the
optical elements.} 
\label{fig:lyot}
\end{figure}

In the Fraunhofer diffraction regime, perfect elimination of on-axis
light can be achieved with a band-limited image mask (Kuchner \&
Traub 2002). Band-limited masks are smooth graded masks that consist
of only a limited range of low-spatial-frequencies. They operate by
placing all of the diffracted light from a point source outside the edges
of the Lyot stop. In the absence of scattered light and aberrations,
the combination of the mask and optimized stop completely removes
on-axis light, creating infinite dynamic range. The light from dim
off-axis sources, such as sub-stellar companions or debris disks,
penetrates the Lyot stop and forms an image at the detector.

Notch filter masks offer the same performance as band-limited masks,
but have more design freedom \cite{kuch03}. By transmitting power at
high-spatial-frequencies, that diffract light well outside of the Lyot
stop opening, notch filter masks afford the construction of sampled graded,
sampled binary, and smooth binary masks. Binary masks are everywhere
either completely opaque or perfectly transmissive.

% This feature mitigates the problem of fabrication-induced phase errors associated
% with masks that change transmissivity.

%, and hence their thickness.

Eighth-order masks are the first group in a large family of
band-limited and notch filter masks, called ``high-order'' masks,
that trade inner-working-angle (IWA) and Lyot stop throughput for
resistance to aberrations (Kuchner, Crepp, \& Ge 2005, hereafter
KCG05). To demonstrate some of these properties in practice, we have built
and tested several eighth-order notch filter masks. The results of
our study are presented with the following: first, we discuss
the advantages of using a high-order image mask for coronagraphy 
($\S$\ref{sec:highorder}); then, we describe the
designs of the masks we have built ($\S$\ref{sec:design}), and
our experimental setup and techniques ($\S$\ref{sec:setup});
finally, we make direct comparisons between the eighth-order masks
and a fourth-order mask, which is used as an experimental control,
by testing their sensitivity to tilt and focus aberrations
($\S$\ref{sec:results}). 

% and highlight the conclusions of our study below
%In $\S$\ref{sec:design}, we discuss the designs of the masks we have
%built.; in $\S$\ref{sec:setup}, we describe our experimental setup
%and techniques; in $\S$\ref{sec:results}, we show results.

\section{HIGH-ORDER IMAGE MASKS}
\label{sec:highorder}

The ``order'' of an image mask describes the intensity transmission
of the mask near the optical axis. High-order masks have a broad central 
opaque region, and transmit less light near the optical axis than lower-order 
masks. To establish some notation, we define the order explicitly as: the exponent of the 
first term in the squared modulus of the expansion of the low-frequency part of the 
mask amplitude transmission, $\hat M(r)$, with respect to distance from the optical 
axis, $r$, where $r << \lambda_{min}/D$. For example, the $\hat M(r)=\mbox{sin}^2(\alpha \, r)$ 
band-limited mask is fourth-order: $|\hat M(\alpha \, r)|^2 \sim (\alpha \, r)^4$, where 
$\alpha$ is a constant that controls the coronagraph's bandwidth, IWA, and Lyot stop throughput. 
Since band-limited masks do not explicitly manipulate the phase of light in theory (which provides
excellent broadband performance), they have symmetric transmission profiles. As a result, the order of 
masks increases by increments of four, starting with fourth-order (see Kuchner 2004). The next 
higher-order mask function in the sequence, an eighth-order mask, is simply a linear combination 
of two fourth-order mask functions weighted such that the quadratic term in the amplitude transmission 
cancels. This forces the quartic term, and subsequent low-order cross-terms, in the intensity 
transmission to zero.  

KCG05 show that the following equations describe the amplitude
transmission of two series of eighth-order band-limited image masks:
\begin{equation}
\hat M_{BL}(r)=N \left[ \frac{3n-1}{3n}-\mbox{sinc}^n (\alpha \, r /
n) + \left( \frac{1}{3n} \right) \mbox{cos} (\alpha \, r) \right],
\label{eq:best_throughput}
\end{equation}
and
\begin{equation}
\hat M_{BL}(r)=N \left[ \frac{l-m}{l} - \mbox{sinc}^l (\alpha \, r /
\, l) + \frac{m}{l} \, \mbox{sinc}^m (\alpha \, r / m) \right],
\label{eq:ringing}
\end{equation}
where $N$ is a normalization constant, and $n$, $m$, and $l$ are
positive integers. For a given IWA, the mask functions in
Equation~\ref{eq:best_throughput} generally yield higher Lyot stop
throughput, while the mask functions in Equation~\ref{eq:ringing}
have less off-axis attenuation. Additionally, there is a trade-off
between the Lyot stop throughput and image-plane off-axis attenuation 
as $n$, $m$, and $l$ increase: large exponents yield lower Lyot stop 
throughput but less `ringing'. We use these exponents to refer to the 
various combinations available for constructing eighth-order masks. For
instance, the $m=1$, $l=2$ mask combines the $1-\mbox{sinc}(\,)$ and
$1-\mbox{sinc}^2(\,)$ fourth-order band-limited functions, and an
$n=3$ mask combines the $1-\mbox{sinc}^3()$ and $1-\mbox{cos}(\,)$
functions. All high-order masks, including notch filter masks - both 
sampled and smooth -, use fourth-order band-limited mask functions as 
a basis set for design.

The next higher-order masks that manipulate the amplitude of starlight and not the 
phase are twelfth-order, sixteenth-order, twentieth-order, and so on. Shaklan \& 
Green 2005 (hereafter SG05) have shown with numerical simulations that the order of 
the mask is directly related to its sensitivity to low-spatial-frequency aberrations: 
higher-order masks significantly reduce light leakage due to low-spatial-frequency
aberrations, and can generate better contrast than lower-order masks at a 
given aberration level.\footnote{This technique of generating broad central 
`nulls' is adopted from interferometry; by adding telescopes to an array and 
forming certain configurations, sets of nulling interferometers, each with a 
$\pi$ phase shift at one arm, can be combined to form larger nulling interferometers 
composed of $N$ dishes \cite{rouan04}.}

%Shaklan \& Green 2005 (hereafter SG05) have shown analytically, using 
%the Zernike polynomials, that the order of the mask directly determines 
%its sensitivity to low-spatial-frequency aberrations. The result is that 
%higher-order masks significantly reduce light leakage due to 
%low-spatial-frequency aberrations, and can generate better contrast than 
%lower-order masks at a given aberration level.

%As we discuss in $\S$\ref{sec:aberrations} however the opposite is
%true when the aberration levels are larger than a certain value.

When compared to fourth-order masks, eighth-order masks relax
pointing requirements by almost an order of magnitude (KCG05). They
relax wavefront error requirements for focus, astigmatism, coma, and
trefoil by two orders of magnitude, and help to reduce leakage due
to spherical aberration and various polarization effects (SG05).
Eighth-order masks can also generate regions of high contrast around
stars of appreciable angular size (KCG05, Crepp \& Ge 2006, in preparation). 
However, increasing the order of the mask requires decreasing the Lyot 
stop throughput, when the IWA is fixed. Choosing an optimal mask depends 
upon the application.

The current mission goals and observing strategy for the TPF-C are
to generate better than $10^{-10}$ contrast at angular separations as
close as $\sim 4 \, \lambda_{max}/D$ from targeted main-sequence F,
G, and K stars within the solar neighborhood. KCG05 and SG05 suggest
that an $m=1$, $l=3$ eighth-order mask can meet these criterion, offering
a good compromise between Lyot stop throughput, off-axis attenuation, and
aberration sensitivity. However, if the mission is altered or expanded to
search more luminous stars, such as sub-giant or red giant stars that have
more distant and extended habitable zones (Lopez, Schneider, \& Danchi
2005), the IWA can be increased. In this situation, a higher-order
mask could be used to relax the overall aberration sensitivity of
the instrument and reduce leakage due to the larger stellar angular
sizes, while maintaining a high Lyot stop throughput. For example, a
twelfth-order linear mask composed of $1-\mbox{sinc}(\,)$, $1-\mbox{sinc}^2(\,)$,
and $1-\mbox{sinc}^8(\,)$ functions can be designed to have an IWA of
$10 \, \lambda_{max}/D$ with a Lyot stop throughput of $\sim 64\%$ and
very little off-axis attenuation, over the entire TPF-C bandwidth.

From the ground, higher-order masks can be used to compensate for systems with
low-order aberrations left uncorrected by the adaptive optics (AO) system, possibly
those introduced by telescope flexure, for example. However, the Lyot stop throughput
is often severely restricted by a telescope central obstruction, placing limits on the
order of the mask that can be implemented. We calculate that a telescope with a central
obstruction of width $0.36 \,D$, such as the Hale $200''$ at Palomar, can achieve
$\sim 16\%$ Lyot stop throughput when operating at $2.2 \,\mu m$ with an $m=1$, $l=3$
8th-order radial mask with an IWA of $600$ mas, and that the throughput goes to zero once
the bandwidth is increased beyond $\sim 20\%$. With the same parameters, a twelfth-order
radial mask would have zero throughput even with monochromatic light, regardless of the 
composite functions. In practice, the size of the Lyot stop may be reduced further compared 
to these ideal cases.

\section{MASK DESIGN \& FABRICATION}
\label{sec:design}

We have manufactured four binary notch filter image masks using
$e$-beam lithography: one fourth-order mask and three eighth-order
masks. These masks represent a second generation of technology
development, where we have improved upon the prototype mask
presented in Debes et al. 2004. In the following, we briefly
describe our design strategy and nanofabrication techniques.

% scratch widths $\leq 20 \, \mu$m, and dig diameters $\leq 100 \, \mu$m []

\begin{figure}[!th]
\centerline{
\includegraphics[height=1.6in]{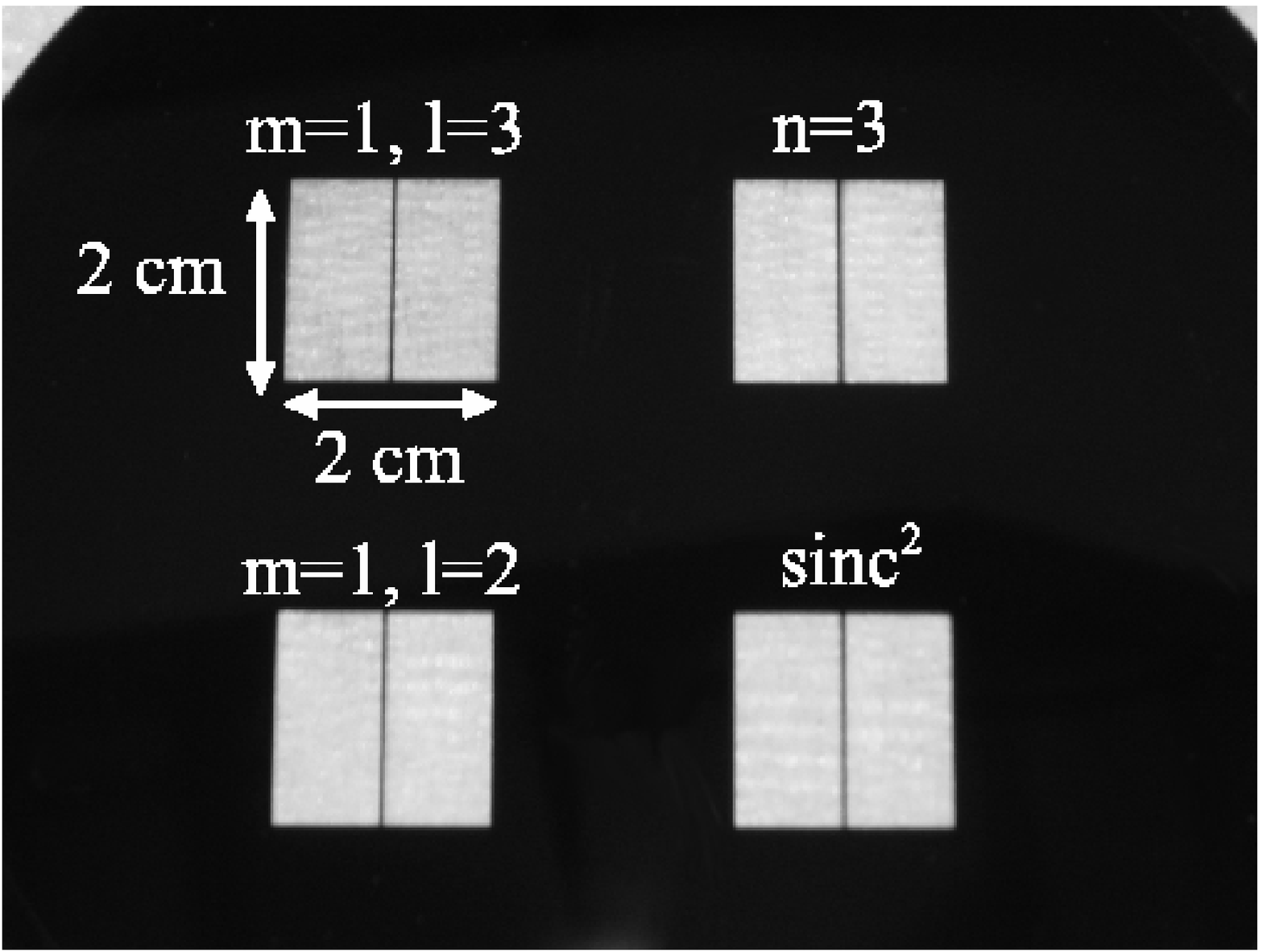}
\hfill
\includegraphics[height=1.6in]{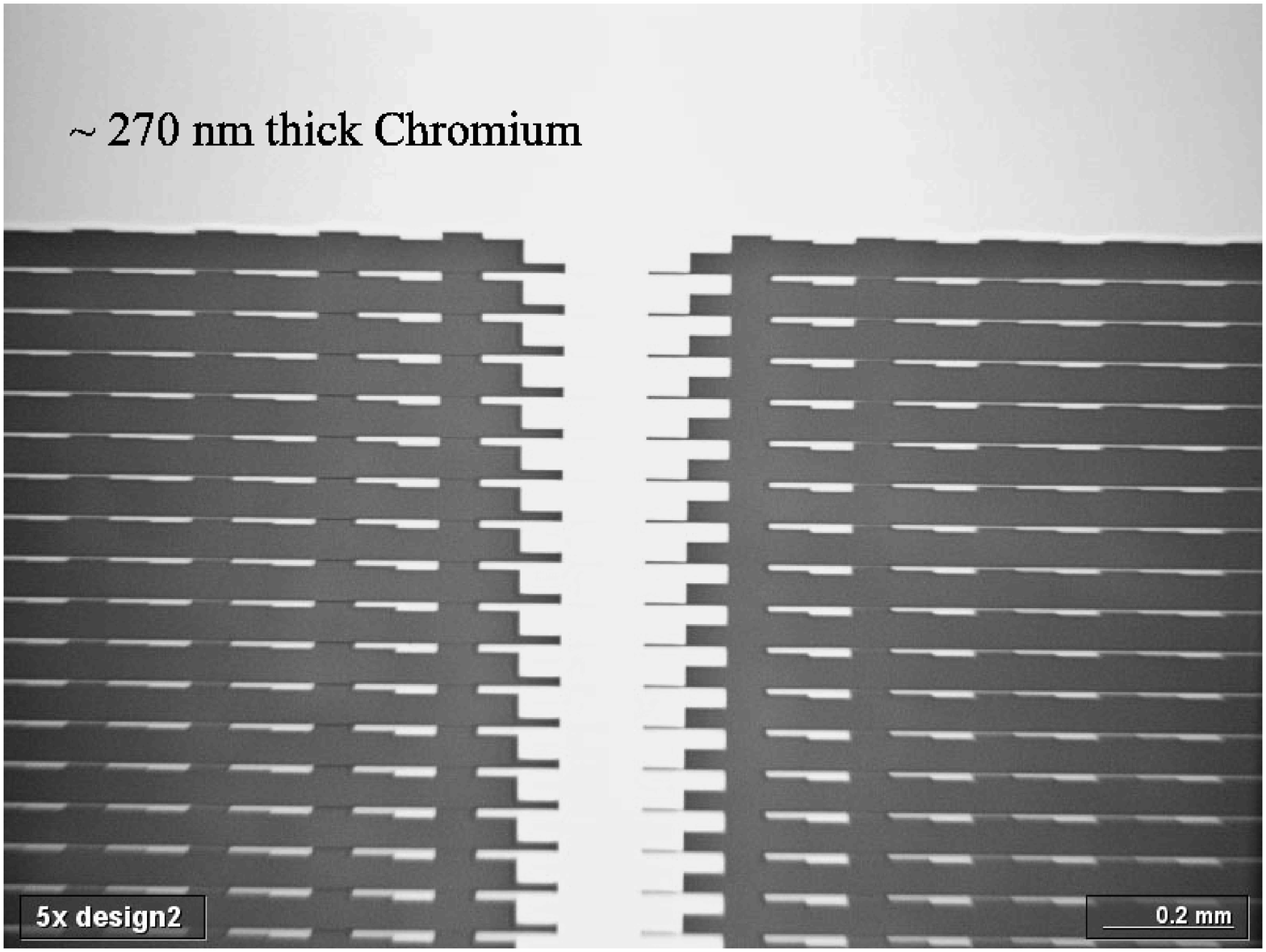}
\hfill
\includegraphics[height=1.6in]{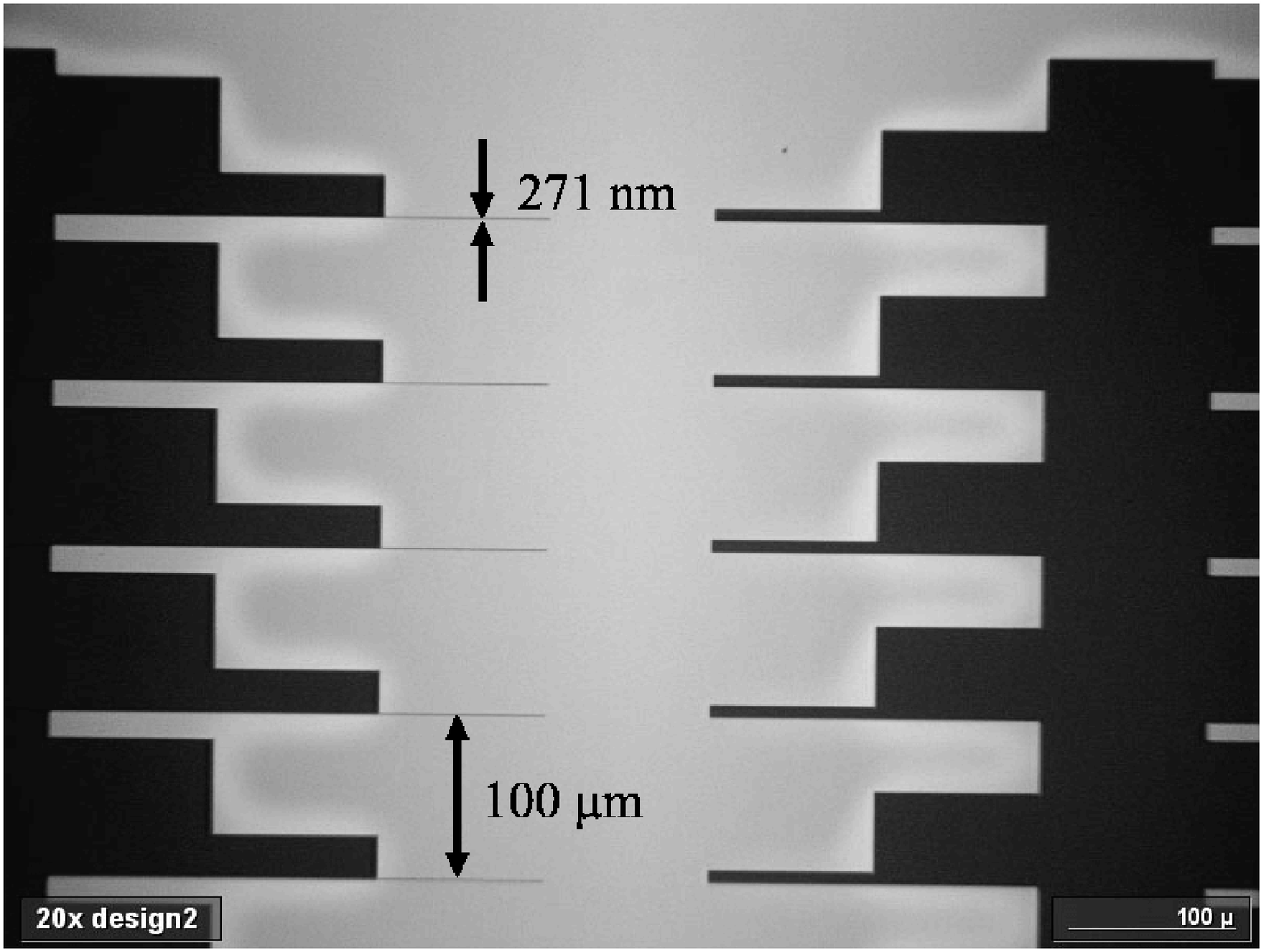}
} \caption{The four linear binary notch filer image masks (left), and
optical microscope false color images of the $m=1$, $l=2$
eighth-order mask at 5x magnification (middle) and 20x magnification
(right). The dark areas in the microscope images are transparent;
this is where the Chrome has been etched away. The spacing between
stripes and the spacing between samples is $\lambda_{min} \, f/ \# =
100 \, \mu m$.} \label{fig:masks}
\end{figure}

The base structure used to mechanically support the opaque portions
of the masks is a 0.7 mm thick piece of Boroaluminosilicate glass with 
a scratch/dig of 20/10. A 270 nm thick layer of Chromium serves as the 
on-axis occulting material, and was deposited onto one side of the glass 
using a Semicore $e$-gun evaporator. Small structures were then dry-etched 
from the Chrome layer with an applied materials cluster tool using a high 
density decoupled plasma composed of Argon, Chlorine, and Oxygen. No 
anti-reflection coating was applied. Figure~\ref{fig:masks} shows a photo 
of the substrate containing all of the designs.

% The manufacturer removed a portion of glass from the bottom of the
% initially round substrate, making a flat edge to facilitate the
% fabrication of orthogonal features and to help with optical alignment
% and mounting.

% eventually conforming to the four different notch filter mask patterns.

Each mask is designed for an $f$/163 or slower beam with a 40 nm
bandwidth centered on the $\lambda=632.8$ nm HeNe laser source. The
focal ratio of the system is large to facilitate the fabrication of
small features in the masks. The physical size or extent
of the masks are 2 cm to a side. Although truncation sets an
outer-working-angle and degrades contrast, notch filter masks can
easily be manufactured large enough to ensure that these effects do
not place significant constraints on the search area and are not the
dominant source of error. Notch filter masks can be designed to have an
azimuthally symmetric search area; however, we have chosen to make linear
masks so that the effective opacity changes in only one direction. This
property simplified the testing of their response to pointing errors.
The FWHM of the image masks (i.e. $2 \times$ IWA) were designed to be roughly 
equivalent such that fair comparisons of their performance could be made 
(Table~\ref{tab:parameters}). Aime 2005 suggests that the mask equivalent-width 
serves as a better proxy for making such comparisons; the authors note that 
the FWHM value differs from the equivalent-width value by $\lesssim 1\%$ for 
each of the individual masks presented here.

Linear binary masks consist of vertically repeating parallel stripes, where 
band-limited or notch filter functions describe the curves in each stripe
\cite{kuch03}. We have made binary masks using the notch filter functions, 
because sampling makes the intricate features near the optical axis a part of 
the design, and not the result of the finite resolution of the lithography
machine. In general, the sampling is not symmetric (Fig.~\ref{fig:masks}).
The low-frequency amplitude transmission of the eighth-order masks follow 
Equations~\ref{eq:best_throughput} and~\ref{eq:ringing}, where $r$ is 
replaced with $x$. 

%In Equations~\ref{equ:best_throughput} and~\ref{equ:ringing}, this
%corresponds simply to replacing $r$

The smallest features in a mask are often found near the optical axis. Their size
depends upon both the IWA and bandwidth, among other parameters. Generally, as the
IWA improves (i.e. gets smaller), the size of the smallest features in the mask
increases, and as the bandwidth widens, the size of the smallest features decreases.
If the IWA is too large or the bandwidth too wide, the smallest features may be too
small to build.

% may ultimately prevent binary masks from generating contrast levels better than $10^{-10}$ with broadband light.

An additional constraint is that the minimum feature sizes should not be smaller than the thickness
of the opaque material. This helps to minimize the waveguide effects associated with binary masks,
and other vector electromagnetic effects that can degrade contrast, especially with broadband light.
Lay et al. 2005 describe some of these potential limiting factors for the TPF-C mission, and suggest
several alternatives for compensation; one of which includes dramatically increasing the focal ratio 
at the mask ($>$ $f$/60), as was done in this experiment. This design strategy was also implemented 
in the Debes et al. 2004 experiment for similar reasons.  

%introduced by the finite thickness of the mask.
%boundaries of the transparent regions in the mask prevents the small
%transparent regions in the mask from acting as waveguides, and producing unwanted
%diffraction (Lay et al. 2005).

%Consequently, our masks have two different IWAs. In each case, the inner-most IWA
%is better than the IWA of a band-limited mask equivalent - that is, a band-limited
%mask with the same Lyot stop throughput\footnote{It is even possible to design
%sampled linear masks such that the {\it average} IWA is better than a band-limited
%mask equivalent.}. This design technique can be implemented on systems that use
%deformable mirrors for wavefront correction, where the contrast on one side of the
%image is improved in a certain region. The mask can be rotated such that the better
%IWA always resides on the side of the image where this ``dark hole" is created
%(see Trauger et al. 2004). NO!

%State-of-the-art deformable mirrors have been shown to improve
%contrast by $4-5$ orders of magnitude over a narrow bandwidth, by
%correcting for scattered light in speckle dominated systems
%\cite{trau04}.

%At low spatial frequencies, sampled binary masks only approximate band-limited masks,
%whereas smooth binary masks are equivalent to band-limited masks in this limit (NO!).

% Often masks are designed for a particular IWA.
% These competing effects

Minimum feature size requirements - both practical and theoretical - limited our ability
to make eighth-order masks with high Lyot stop throughput. In theory, eighth-order masks
can achieve $\sim 50 \%$ Lyot stop throughput with IWAs of $\sim 4 \, \lambda_{max} / D$.
Our masks were {\it designed} however to achieve only $\sim 20\%$ throughput at most,
because we were restricted to making features larger than the thickness of the
Chrome; the smallest feature size in each of the eighth-order masks
is $\sim 270$ nm, whereas the smallest feature size in the
fourth-order mask is 7119 nm. To increase the throughput, we would
have to increase the $f/ \#$, decrease the bandwidth, or decrease
the thickness of the Chrome. Clearly the focal ratio is already
large and the bandwidth is already narrow. Also, we show in
$\S$\ref{sec:results} that increasing the thickness of the Chrome is
the most notable improvement we have made upon the mask presented in
Debes et al. 2004. Thus, future mask development will necessarily
involve using a material that is intrinsically more opaque at visible
wavelengths, such as Aluminum (Semaltianos 2001; Lay et al. 2005). This
will enable the fabrication of masks that have more opacity and smaller
minimum features sizes.

% mention here switching to smooth masks as well as polarization?

The equations describing the exact structure of linear, binary,
sampled eighth-order notch filter image masks are derived in KCG05 .
Table~\ref{tab:parameters} displays the relevant quantities for our
designs, using the same notation (except in KCG05 $f/\# = f$). For each
mask, sampling began at a horizontal distance of
$\, x=\zeta_0 \, \lambda_{min} \, f/\# \,$ from the optical axis.

\begin{table}[!ht]
\centerline{
\begin{tabular}{ccccc}
\hline
\hline
Mask & $\;\;$ $\mbox{sinc}^2$     $\;\;$ & $\;\;$$n=3$     $\;\;$ & $\;\;$ $m=1$, $l=2$    $\;\;$ & $\;\;$ $m=1$, $l=3$ \\
\hline
order                       &     4th$\tablenotemark{a}$  &       8th       &   8th            &   8th        \\
$\mbox{IWA} / (\lambda_{max} / D)$  &   2.350  &   2.372    &   2.332  &   2.356 \\
$\epsilon$                  &   0.488      &       0.674     &     0.716        &   0.759      \\
$N$                         &   1.01444077 &      0.99098830 &   1.93640370     &   1.47114548 \\
$\hat M_{0_A}$              &   0.01423518 &      0.07833984 &   0.05989364     &   0.06702707 \\
$\hat M_{0_B}$              &   ---        &      0.01803281 &   0.03028228     &   0.02278117 \\
$C$                         &   ---        &     -0.25123206 &  -0.51959437     &  -0.35644301 \\
$\zeta_0$                   &   0.28800972 &      0.25875213 &   0.25877876     &   0.25874218 \\
Smallest Feature            &   7119 nm    &      271 nm     &    270 nm        &   270 nm     \\
Theoretical Throughput      &   40.4\%     &       21.2\%    &    17.4\%        &   13.7\%   \\
Experimental Throughput     &   30.7\%     &       15.7\%    &    13.0\%        &   10.4\%   \\
\hline
\end{tabular}
} \caption{Mask design parameters for a bandpass of $632.8 \pm 20$
nm. For an elliptical or circular primary mirror, the Lyot stop
throughput of a linear mask is given by: $T=1-\frac{2}{\pi}[\epsilon
\sqrt{1-\epsilon^2}+\mbox{arcsin}(\epsilon)]$, where $0 \leq
\epsilon \leq 1$ is a dimensionless parameter that controls the
width of the zones of diffracted starlight at the edges of the Lyot
stop. We were able to achieve $\sim 75\%$ of the theoretical Lyot
stop throughput for each mask in the experiment.}
\tablenotetext{a}{Fourth-order masks have only one $\hat M_0$ and no
$C$ (KCG05).} \label{tab:parameters}
\end{table}

%IWA bandlimited=2.356 lambda max!

\section{EXPERIMENTAL SETUP}
\label{sec:setup}

The design of the University of Florida coronagraphic testbed is
that of a standard transmissive Lyot coronagraph without wavefront
correction, as depicted in Figure~\ref{fig:lyot}, and similar to the
setup described in Debes et al. 2004. ``Starlight'', generated by a $\lambda=632.8$ nm 
HeNe laser for monochromatic testing, passes first through a set of neutral density (ND) 
filters and is then focused by a microscope objective lens into a $4 \, \mu$m single-mode fiber 
for spatial filtering. The fiber exit-tip serves as a bright point source. This expanding beam, 
N.A.$\,=0.12$, is collimated and then truncated by a circular $\sim 3$ mm diameter iris, simulating the 
primary mirror of an off-axis telescope. Optics downstream from the iris are high quality achromat 
doublets capable of handling future broadband tests. The first achromat, $f=500$ mm,
focuses the light onto the substrate containing all of the notch
filter image masks. The substrate is mounted onto a precision x-y-z
stage for fine adjustments. The light is then re-collimated by an
identical achromat. In the Lyot plane, an optimized Lyot stop blocks
the light diffracted by the mask at the location of the reimaged
entrance pupil. The Lyot stop size is adjustable and takes the shape
of the intersection of two overlapping circles, since the masks are
linear. The remaining light is then focused onto an SBIG ST-2000XM
CCD detector where images are taken for analysis. The images are sampled 
approximately $10 \times$ more frequently than the Nyquist frequency with 
$7.4 \, \mu m$ pixels. The anolog-to-digital converter has a bit depth of 16, 
and the CCD has a factory quoted RMS read noise of 7.9 $e-$. These sources of 
noise are always at least two orders of magnitude smaller than the measured 
contrast values when used in concert with our experimental techniques, which 
are described in the following.

%The setup does not include vibration isolation, thermal control, or a vacuum chamber.
%The experiment uses off-the-shelf $\, \lambda/20 \,$ optics.
% mounted onto a $5'' \times 10''$ optical bench.
% 10 micron increment x-y-z stage for fine adjustment alignment.. may want to put in result section

%We have found it useful to include in the testbed a simulated planet of variable
%brightness and position. This can easily be added with a series of beam splitters,
%folding mirrors, and neutral density (ND) filters. The planet serves as an intuitive tool,
%and can be used to direclty measure the IWA and off-axis throughput of the coronagraph;
%although, the beam-splitters add unnecessary scattered light to the system. Shouldn't I
%put a picture in of the planet then!?

\begin{figure}[!ht]
\centerline{
\includegraphics[height=1.7in]{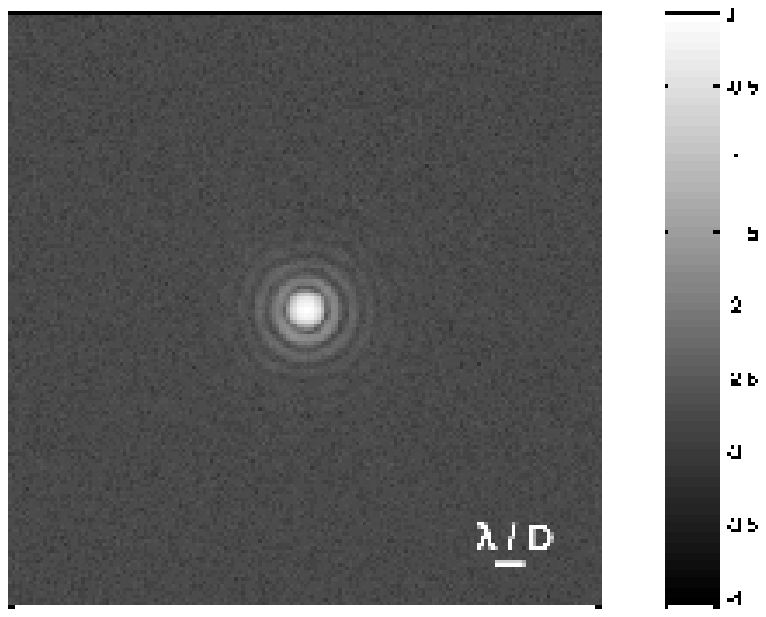}
\hfill
\includegraphics[height=1.7in]{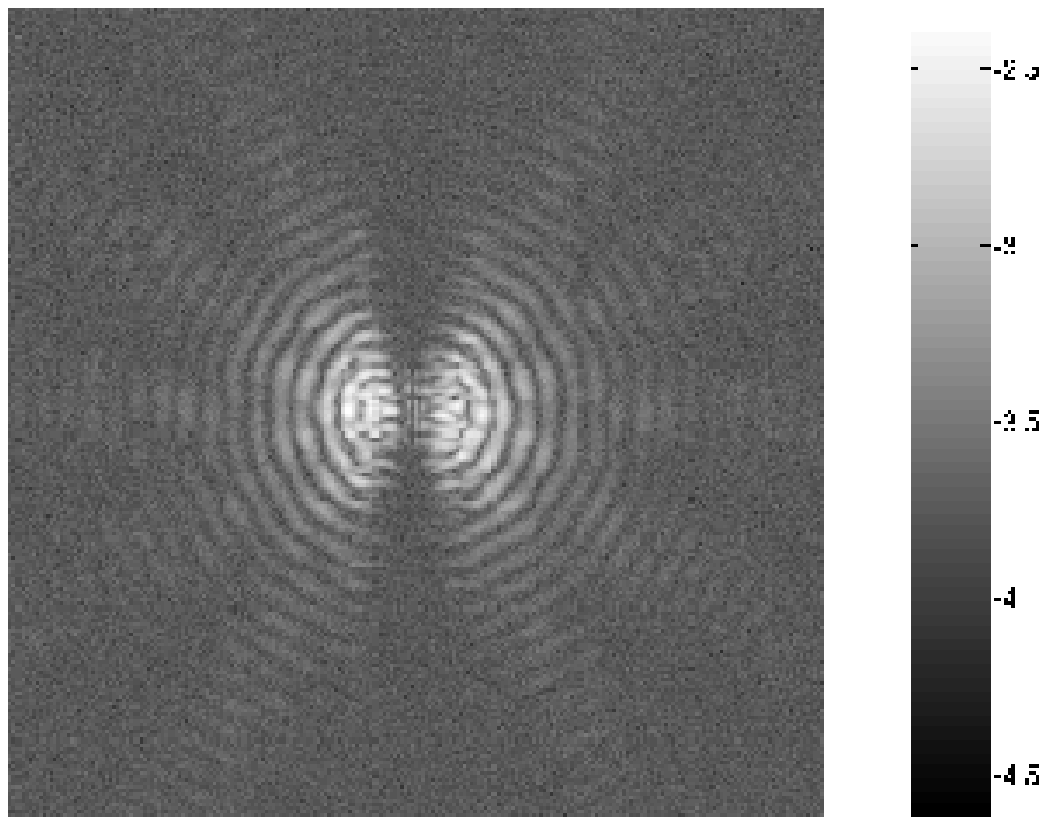}
\hfill
\includegraphics[height=1.7in]{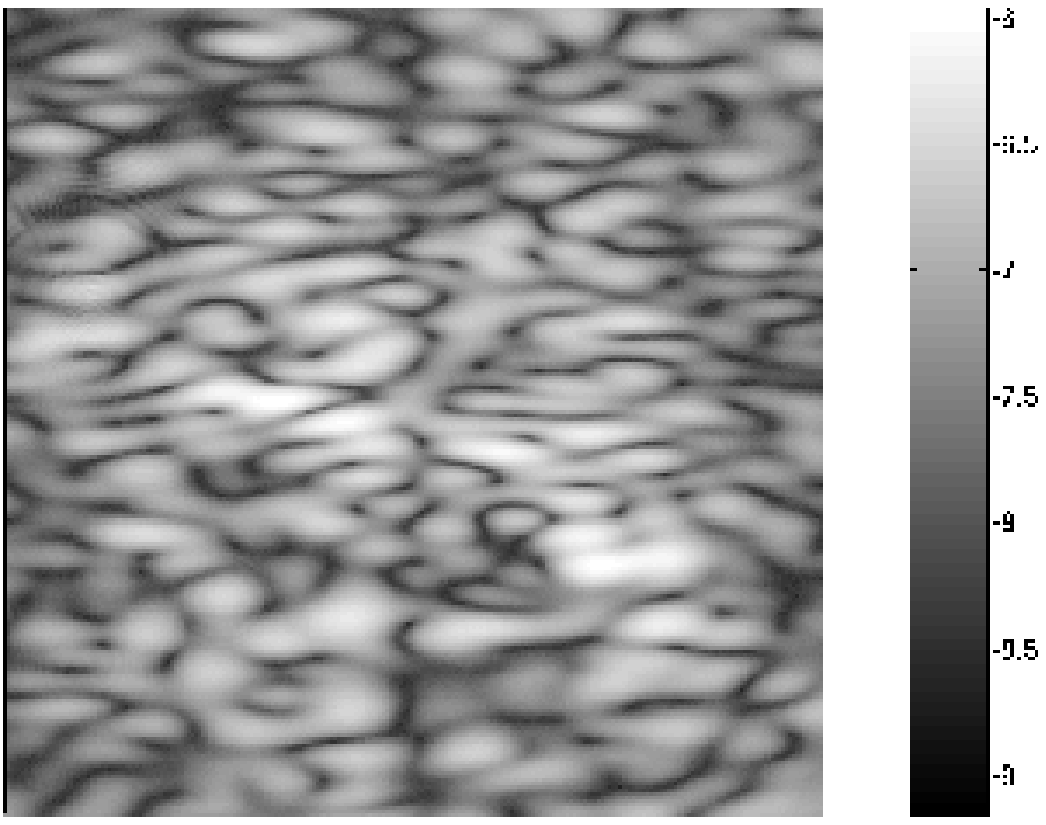}
} \caption{Laboratory images of the simulated star without the coronagraph
(left), the $m=1$, $l=2$ mask aligned over the star (middle), and the star with both the
mask and Lyot stop in place (right). Intensities are plotted on a logarithmic
scale. The image of the star shows the angular size scale of the telescope. 
Speckles created by imperfections in the optics limit the dynamic range of the 
coronagraph creating a noise floor at the $\sim 10^{-7}$ level near the IWA. The contrast is
calculated by dividing the relative intensity, shown in the image on the right and 
later in Fig.~\ref{fig:rel_int}, by the Lyot stop throughput and mask intensity 
transmission; this accounts for the off-axis attenuation of the coronagraph.}
\label{fig:star}
\end{figure}

In order to measure contrast, we perform a comparison of images
taken with and without the coronagraph in place. Due to the extreme
contrast levels involved, we use a combination of the ND filters and 
the linearity of the CCD to calculate relative intensities and to generate 
high signal-to-noise ratio images in a reasonable amount of time. We first 
attenuate the laser light with the ND filters, which are placed upstream 
from the single-mode-fiber to ensure that their aberration effects are 
negated, and take an image of the star without the mask and without the Lyot 
stop in the optical train. Then, the mask and Lyot stop are inserted into place, 
and the ND filters are removed. In this final image, the intensity values at each 
pixel are divided by the average flux within the FWHM of the image of the star 
taken without the coronagraph, and normalized to the integration times and
ND filters used. We define the resulting value, at each pixel, as the relative 
intensity. The contrast then is simply the relative intensity divided by the Lyot 
stop throughput and band-limited (i.e. low-frequency) part of the mask intensity 
transmission at that position. Figure~\ref{fig:star} shows images of the star at 
various steps in the procedure. We could have chosen a less conservative definition of 
contrast by instead normalizing to the interpolated peak intensity of the imaged 
source, rather than the average flux within the FWHM (SG05); however, these two 
values differ by less than a factor of two. It is not necessary to add and remove 
the Lyot stop in order to measure contrast, but we find that doing so facilitates 
calculation of the off-axis throughput. 

%We find that the transmission values of the ND filters quoted by the manufacturer are accurate
%to better than $~20\%$. This uncertainty yields small relative error bars in contrast when performing such
%large dynamical range tests.

% The middle and right images display the multiplicative factor relating
% the integration times and neutral density filters used to accentuate the
% structure in the PSF.

Linear binary masks approximate band-limited masks only when the
vertical angular size of the stripes in the mask are smaller than
$\lambda_{min}/D$, the resolution of the telescope. To ensure that
the masks diffract light appropriately, we increased the focal ratio
of the system in the first image plane from the initial design of
$f/\# \,=163$ to $f/\# \, \approx 187$, by shrinking the size of the
entrance aperture. This also resulted in an improvement of the
masks' effective IWA by the same factor. The Lyot stop size was set
conservatively to obtain $\sim 75\%$ of the maximum theoretical
throughput, so that small mis-alignments did not result in
diffracted light leakage through the center of the stop onto the
detector. Table~\ref{tab:parameters} shows the designed IWAs before
increasing the focal ratio, and the experimental Lyot stop
throughput achieved.

%\begin{figure}[!ht]
%\centerline{
%\includegraphics[height=2.2in]{iris_w_scale_magick_gimp_400.eps}
%\hfill
%\includegraphics[height=2.2in]{lyot_stop_newest_gimp_400.eps}
%}
%\caption{Optical microscope images of the inner-edges of the entrance iris (left)
%and optimized Lyot stop (right). ...}
%\label{fig:microscope_stops}
%\end{figure}

\section{RESULTS}
\label{sec:results}

\subsection{Chrome Transmission \& Relative Intensities}
\label{sec:chrome}

\begin{figure}[!ht]
\centerline{
\includegraphics[height=2.2in]{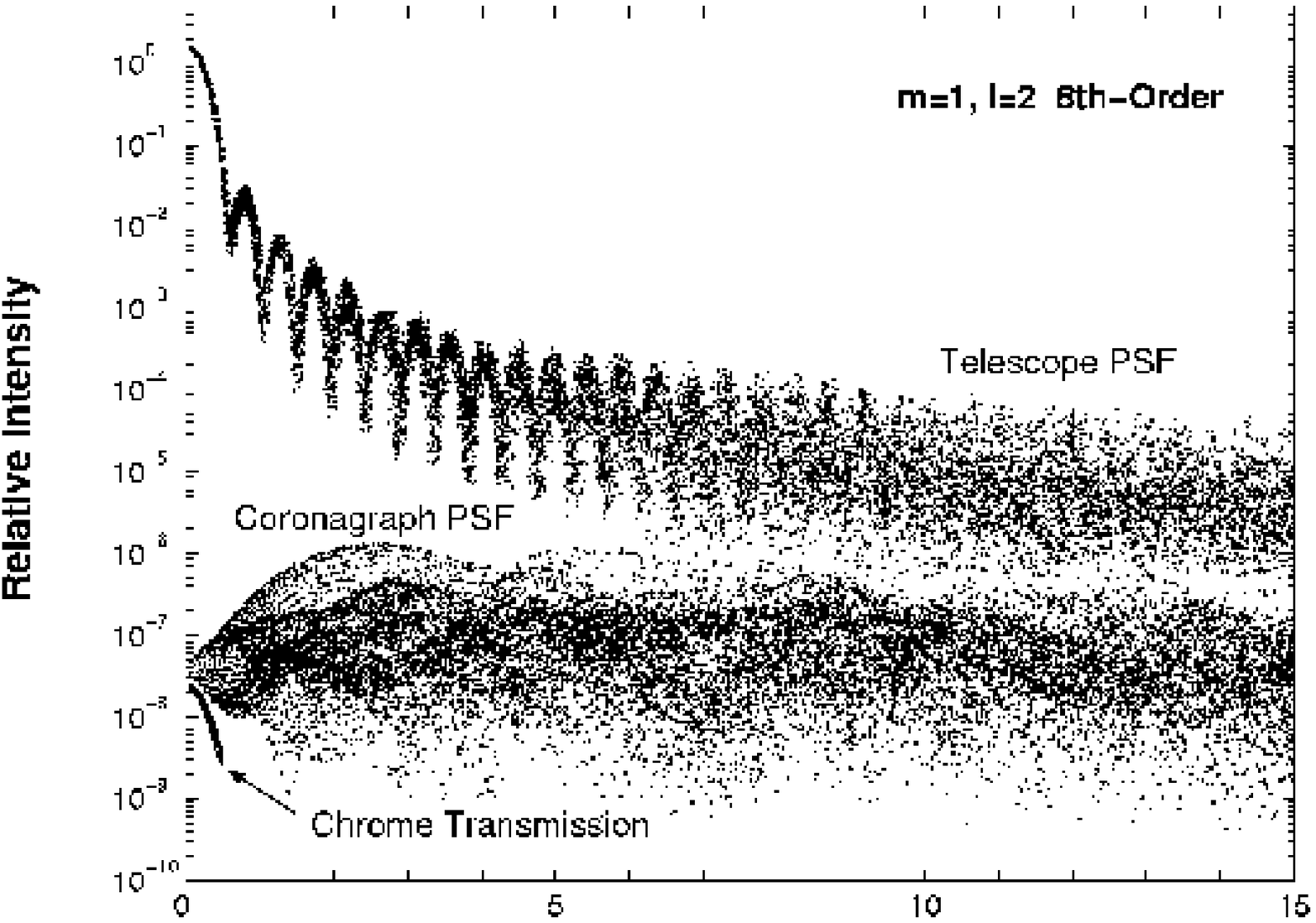}
\hfill
\includegraphics[height=2.2in]{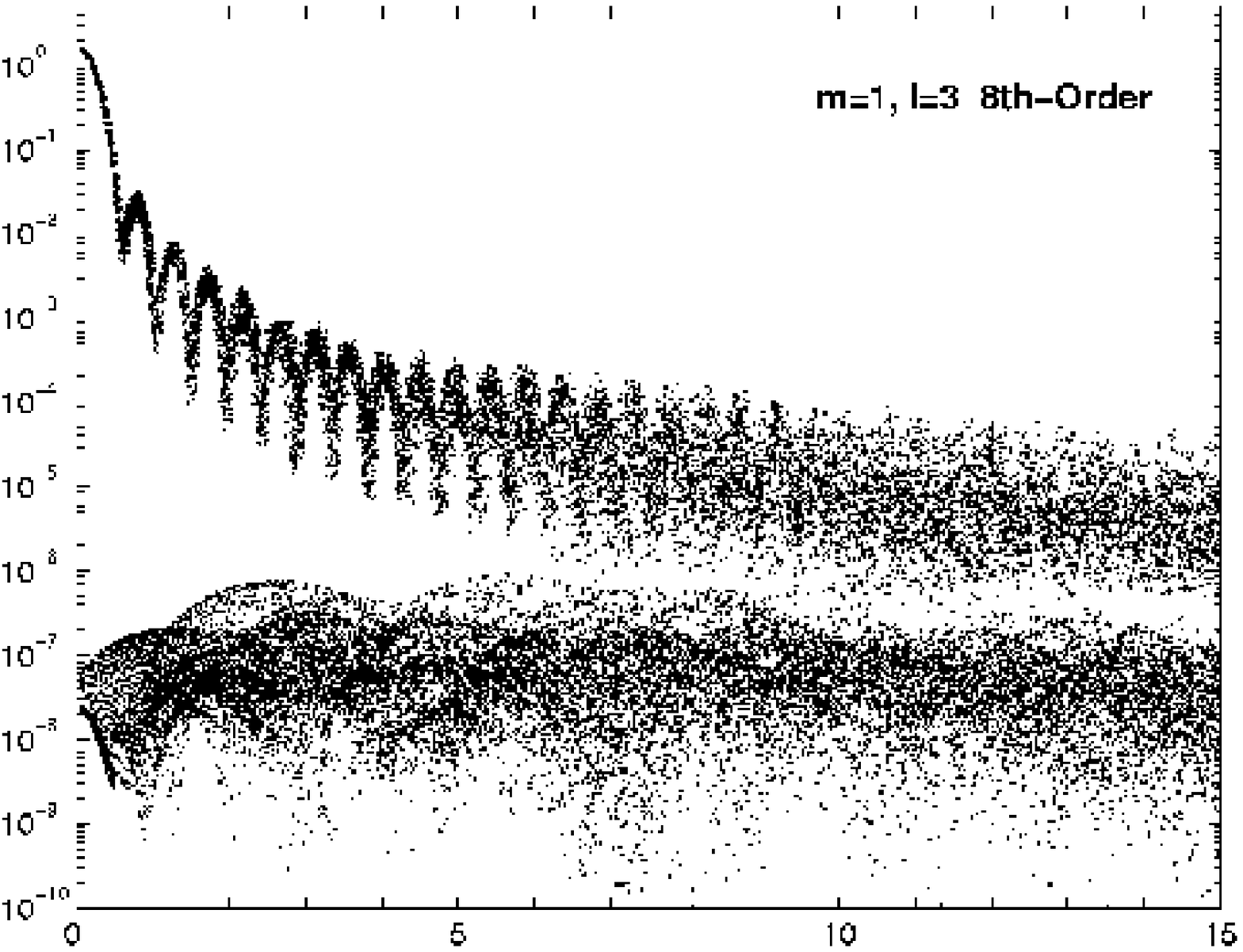}
} \centerline{
\includegraphics[height=2.2in,width=3.12in]{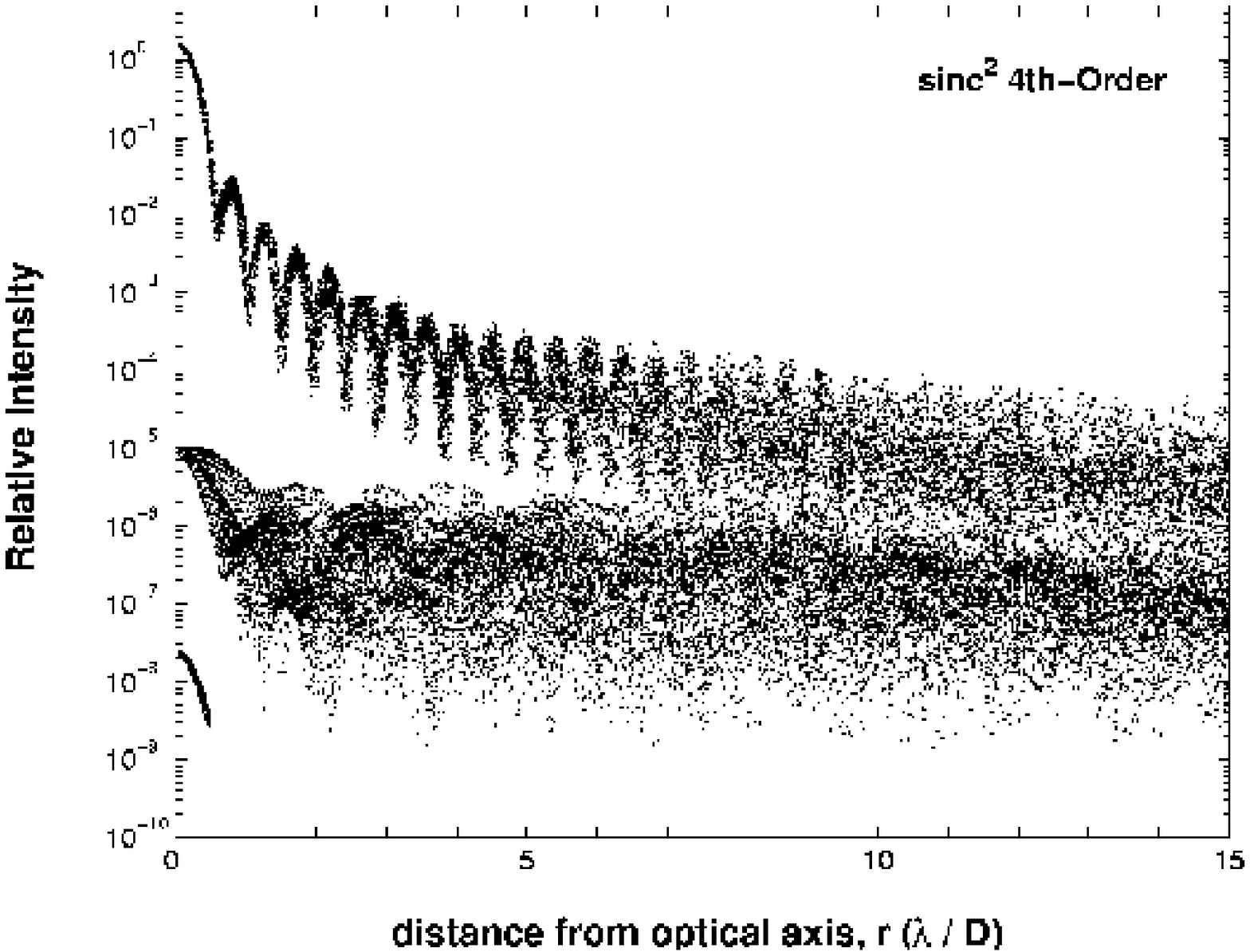}
\hfill
\includegraphics[height=2.2in,width=2.87in]{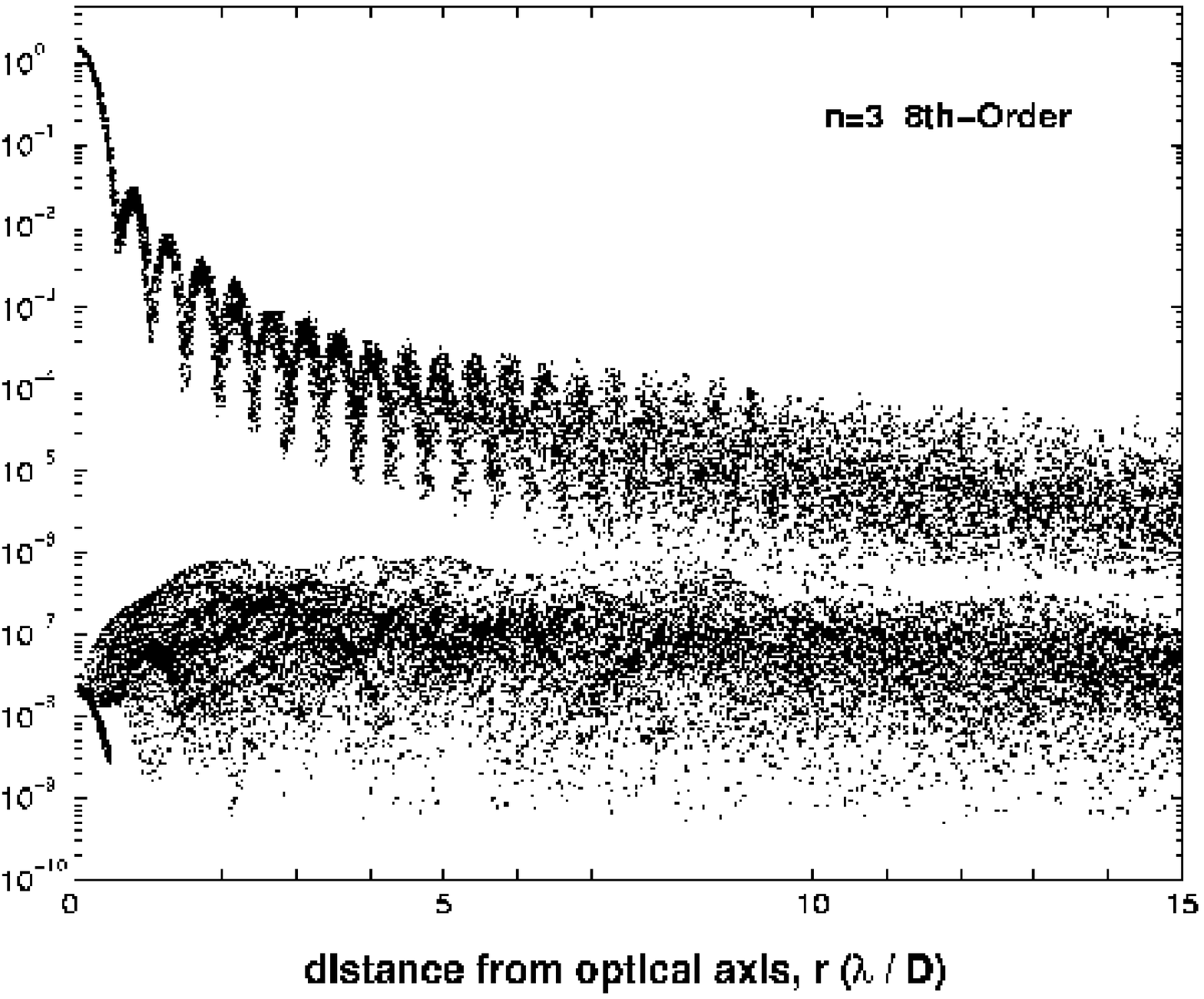}
} \caption{Telescope PSF, coronagraph PSF, and Chrome transmission
for each mask using a circular entrance aperture. The Chrome
transmission was measured at the center of the substrate. Inside the
IWA, the eighth-order masks block more scattered light and reduce
the effects of low-order aberrations. The thickness of the Chromium
does not limit the performance of the coronagraph, but does transmit
a non-negligible amount of light which contributes to the noise floor.
Contrast is calculated by dividing the coronagraph PSF by the intensity
transmission of the mask and the Lyot stop throughput.}
\label{fig:rel_int}
\end{figure}

The performance of the prototype mask presented in Debes et al. 2004
was limited by the transmission of light directly through the Chromium
occulting layer. In this study, we have increased the thickness of
this layer from 105 nm to 270 nm.  Using the transmission curve in
Debes et al. 2004, we calculate that the peak transmission should
improve from $7.5 \times 10^{-4}$ to $9.2 \times 10^{-9}$. We
measure a peak transmission of $2.3 \times 10^{-8}$. This is slightly 
worse than predicted, but opaque enough for this application nevertheless 
(Fig.~\ref{fig:rel_int}). The discrepency in these values can be 
understood by considering inhomogenieties in the thickness of the Chromium 
layers. A $< 5$ nm deviation in each makes up the difference. The contribution 
of transmission directly through the Chromium to the limiting contrast in this experiment
is approximately one order of magnitude smaller than the scattered light floor near
the IWA, and less so in the regions where the masks have little off-axis 
attenuation. A more demanding application, such as the TPF-C, would, of course, 
require a material with a higher opacity.

%Figure~\ref{fig:rel_int} compares the transmission of the Chrome, at
%the center of the substrate, to the relative intensities achieved
%with each of the masks.

Figure~\ref{fig:rel_int} also shows a difference in the amount of
light transmitted near the optical axis, $r \lesssim \lambda / D$,
between the fourth-order mask and the eighth-order masks. This is
evidence that the eighth-order masks are blocking more scattered
light, simply because they are wider, and reducing the effects of
low-order aberrations (we test this latter claim more carefully in
$\S$\ref{sec:aberrations}). These properties are not seen as clearly
in contrast curves, where, interior to the IWA, the intensity
transmission of the mask controls the detection threshold.

\subsection{Contrast Measurements}
\label{sec:contrast}

In a system dominated by scattered light, the relative intensities
should scale according to the size of the Lyot stop. An equivalent
statement is that in a system dominated by scattered light, the {\it
contrast} should be {\it independent} of the Lyot stop size for a
given mask (assuming, of course, that the Lyot stop is smaller than
the diffracted light pattern in the Lyot plane). We find that the
contrast generated by our masks is independent of the undersizing of
the Lyot stop,\footnote{This is not the case when aberrations are
added ($\S~\ref{sec:aberrations}$).} and that the coronagraph's 
performance is limited by scattered light. The image plane speckles 
seen in Figure~\ref{fig:star} are the result of wavefront distortions,
created by imperfections in the optics. In practice, speckles can
mimic and often overwhelm the light of dim companions.

\begin{figure}[!ht]
\centerline{
\includegraphics[height=3.5in]{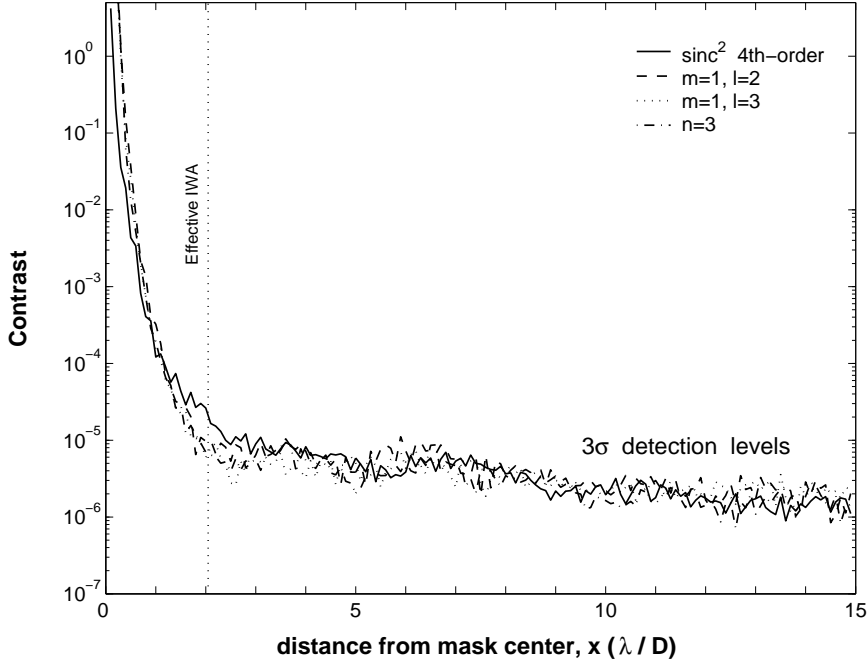}
}
\caption{Experimental $3\, \sigma$ detection limits for each mask. The effective IWAs
were calculated taking into consideration the change in focal ratio from the initial
mask designs.}
\label{fig:contrast}
\end{figure}

We calculated contrast for pixels within a $10 \,\lambda / D \times
30 \, \lambda / D$ section across the center of the PSF by dividing
the relative intensities by the Lyot stop throughput and
band-limited part of the mask intensity transmission, $|\hat
M(x)|^2$, at each position, $x$, the horizontal distance of pixels
from the optical axis. That way, the direction in which the halo of
speckles decreased in flux coincided with the direction of opacity
change in the masks. We were able to achieve $2 \times 10^{-6}$
contrast at $3 \, \lambda / D$ and $6 \times 10^{-7}$ contrast at
$10 \, \lambda / D$, as quoted in the abstract. In
Figure~\ref{fig:contrast}, we plot the corresponding $3 \, \sigma$
detection limits, where $\sigma$ is defined as the standard deviation
of the scattered light noise floor at a given location in the final
image plane. The eighth-order masks reduce the amount of scattered 
light close to the optical axis, and slightly out-perform the fourth-order 
mask near the IWA as a result of their more steeply increasing intensity 
transmission in that region. The contrasts in the rest of the search 
area are essentially identical.

To calculate the Strehl ratio of our system, we compared a model of
the Airy pattern incident onto the mask with experimental data. We
find that the Strehl ratio exceeds $98 \%$. Combining this with the
fact that the coronagraph is speckle dominated and the optical
quality of our achromats are high (see $\S$\ref{sec:aberrations}),
we conclude that the detection limits shown in
Figure~\ref{fig:contrast} represent the approximate deepest contrast
that is achievable from the ground with current AO systems, using
this type of coronagraph.

% $D=2.68 \pm 0.16$ mm
% The final outer-working-angle is $84 \lambda / D$.

% We note here that the results quoted in Debes et al. 2004 are relative intensity divided by Lyot stop
% throughput, and not necessaryily contrast. These two quantities differ by only a factor of approximately
% unity for angular positions outside of the IWA; inside the IWA, they differ by orders of magnitude.

\subsection{Aberration Sensitivity}
\label{sec:aberrations}

\begin{figure}[!ht]
\centerline{
\includegraphics[height=2.7in]{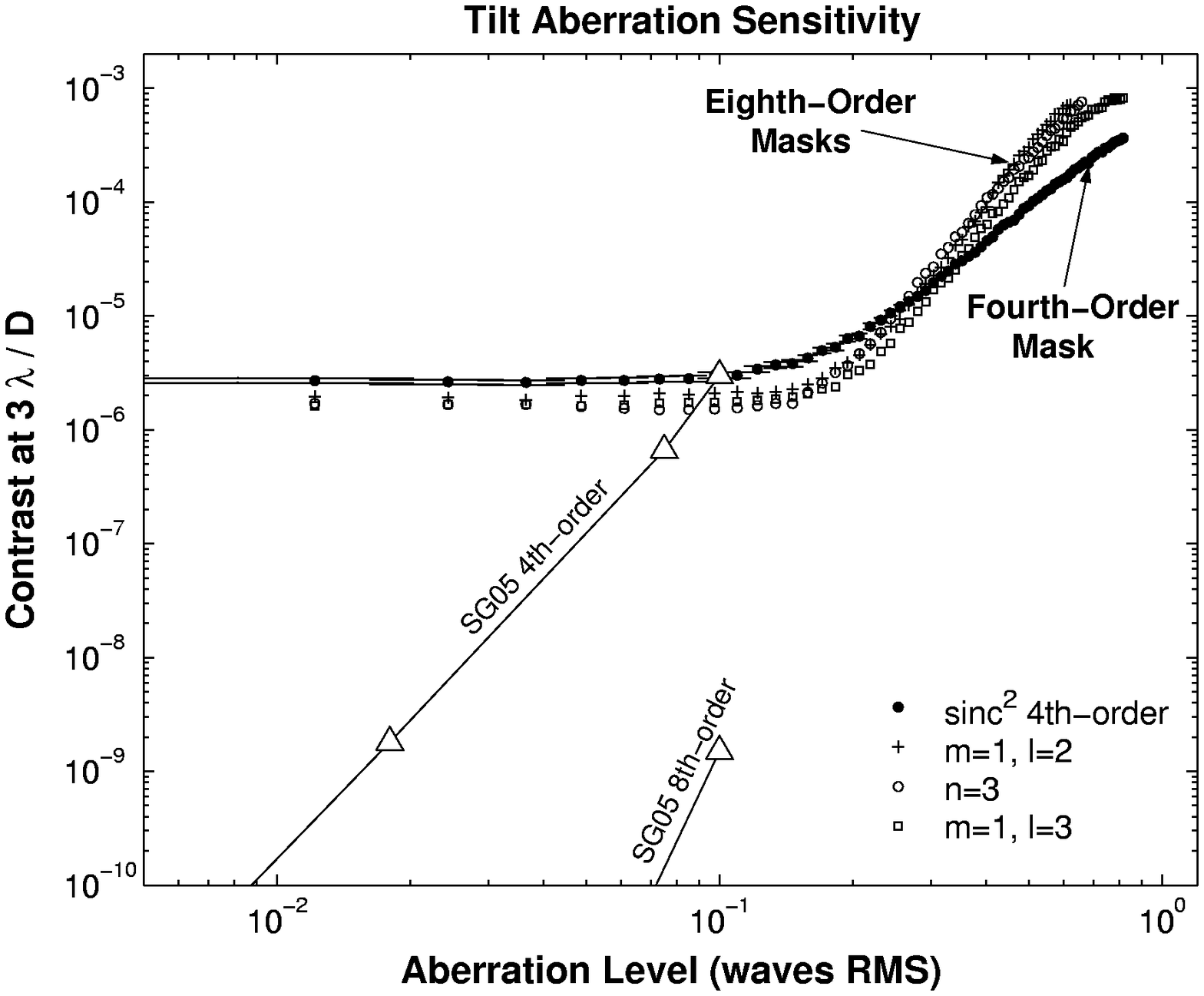}
\hfill
\includegraphics[height=2.7in]{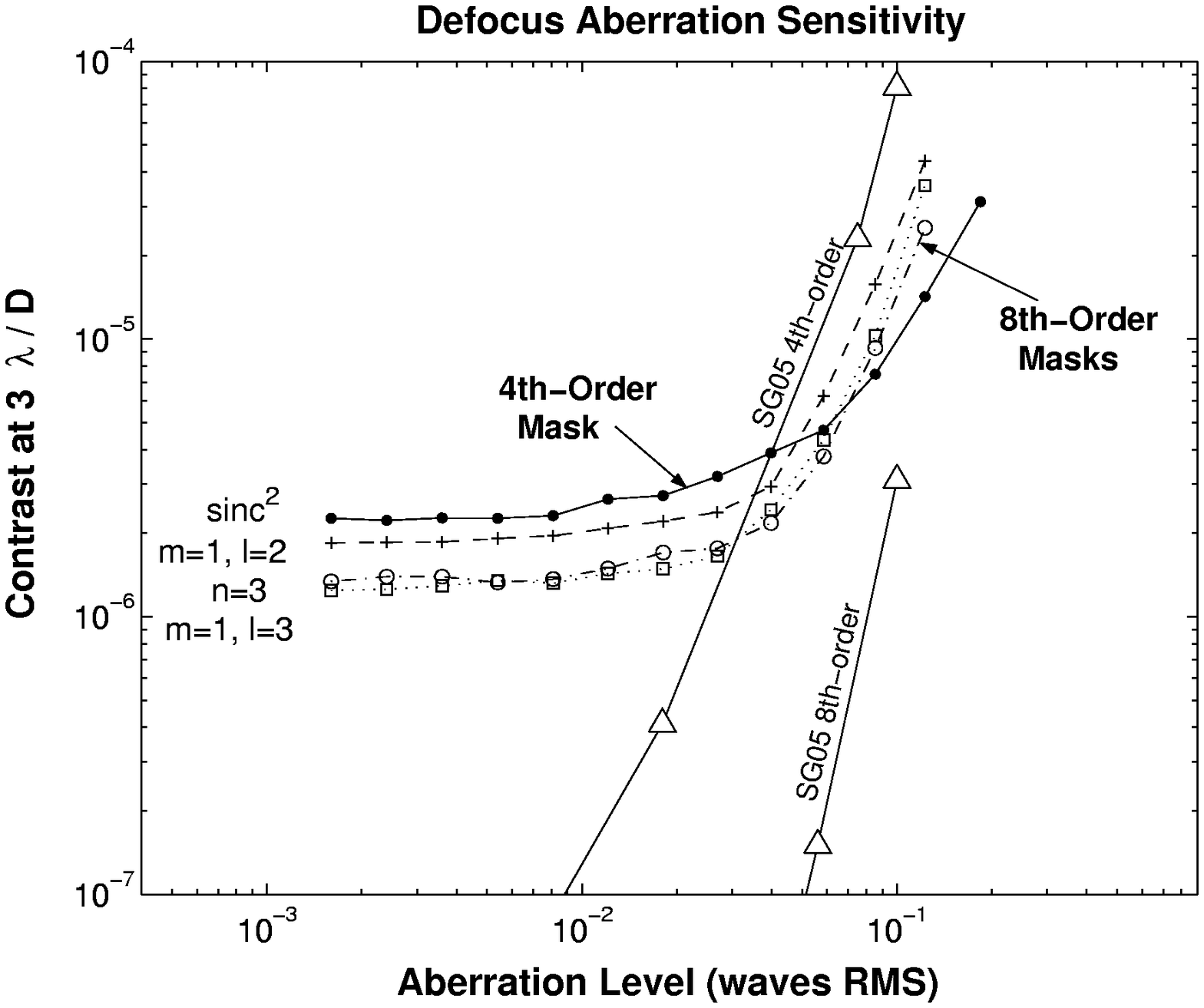}
} \caption{Coronagraph sensitivities to tilt (left) and focus
(right) aberrations for each mask. The theoretical predictions of
SG05 are over-plotted for comparison (see text). Scattered light
prevented measurement of the diffracted light response of the masks
at small aberration levels. Uncertainties in the measurements for the 
fourth-order mask are shown in the tilt graph; the errorbars are on 
the order of the size of the datapoints and representative
for all of the experimental curves shown in both graphs. The focal 
ratio of our system was large enough to warrant tilt realignment for 
each focus datapoint, but not focus realignment for each tilt datapoint; 
this motivated the linear versus logarithmic measurement increments used
for acquiring data.} 
\label{fig:aberr}
\end{figure}

We tested the tilt and focus aberration sensitivities of each mask
by introducing small alignment errors to the substrate in the image
plane. The results are shown in Figure~\ref{fig:aberr}, where we
plot the contrast at $3 \, \lambda / D$ as a function of the
distance that the masks were displaced. The scattered light floor
limits the dynamic range of the coronagraph at small aberration
levels; in this regime, the masks generate contrasts to within a
factor of two of one another, as shown in the previous section
(Fig.~\ref{fig:contrast}). At large aberration levels, where
diffracted light dictates the contrast, there is a clear dichotomy
in the masks' behavior.

%Linear masks are completely immune to pointing errors in the
%orthogonal direction, where $x=0$.

%Linear masks are completely immune to misalignments along the
%axis of the mask, i.e. tip aberrations, and may be used to search
%binary systems for dim companions.

%To avoid counting pixels directly behind the mask, only those pixels
%opposite the side where the mask was displaced were used.

% For intuitive reasons, we plot the tilt dependence versus
% $f \lambda / D$; a displacement in the first image
% plane corresponds to an angular separation on the sky.

To measure the tilt aberration sensitivity of each mask, we moved
the substrate laterally across the center of the image of the star
in 5 $\mu $m increments. We find that the eighth-order masks are
easier to point than the fourth-order mask. To quantify this
statement, we calculated the width of the pointing ``sweet-spot''
for each mask, where the contrast at $3 \, \lambda / D$ is flat to
within $2 \, \sigma$ and limited by scattered light. The mean width
of this zone for the eighth-order masks is $1.06 \pm 0.03 \, \lambda
/ D$; the width of this zone for the fourth-order mask is $0.20 \pm
0.05 \, \lambda/D$, a factor of $\sim 5$ smaller. With an 8 m
telescope operating at $\lambda = 0.5 \, \mu m$, these tolerances
correspond to pointing accuracies of $6.8$ mas and $1.3$ mas
respectively. (Lloyd \& Sivaramakrishnan 2005 discuss tip/tilt errors 
in Lyot coronagraphs in detail. For a practical application, see 
the description of the AEOS coronagraph by Lloyd et al. 2001.)

To measure the focus aberration sensitivity of each mask, we moved
the substrate along the optical axis in logarithmic increments of
0.18 dex. With a similar analysis, we find that the eighth-order
masks are also less susceptible than the fourth-order mask to focal
misalignments. The eighth-order masks provide the same contrast as
the fourth-order mask in a system with $\sim 4$ times as large an
RMS wavefront error.

These results (Fig.~\ref{fig:aberr}) depend upon the 
amount of scattered light present in the system. If the scattered light 
levels were reduced, it would be possible to measure the response of the 
masks to smaller aberrations; the width of the tilt and focus sweet-spots 
would decrease as the contrast improves, and the relaxation ratios would
change. KCG05 estimate that eighth-order masks should relax pointing
requirements relative to fourth-order masks by a factor of $\sim 6$
in a system with no scattered light designed to achieve $10^{-10}$
contrast at $3 \, \lambda / D$. SG05 have performed more careful
calculations and predict an even larger relaxation ratio of 16 in
the allowable tilt RMS wavefront deviation, when comparing
eighth-order masks to fourth-order masks in an ideal system designed
to achieve $10^{-12}$ contrast at $4\,\lambda/D$.

\begin{figure}[!ht]
\centerline{
\includegraphics[height=2.1in]{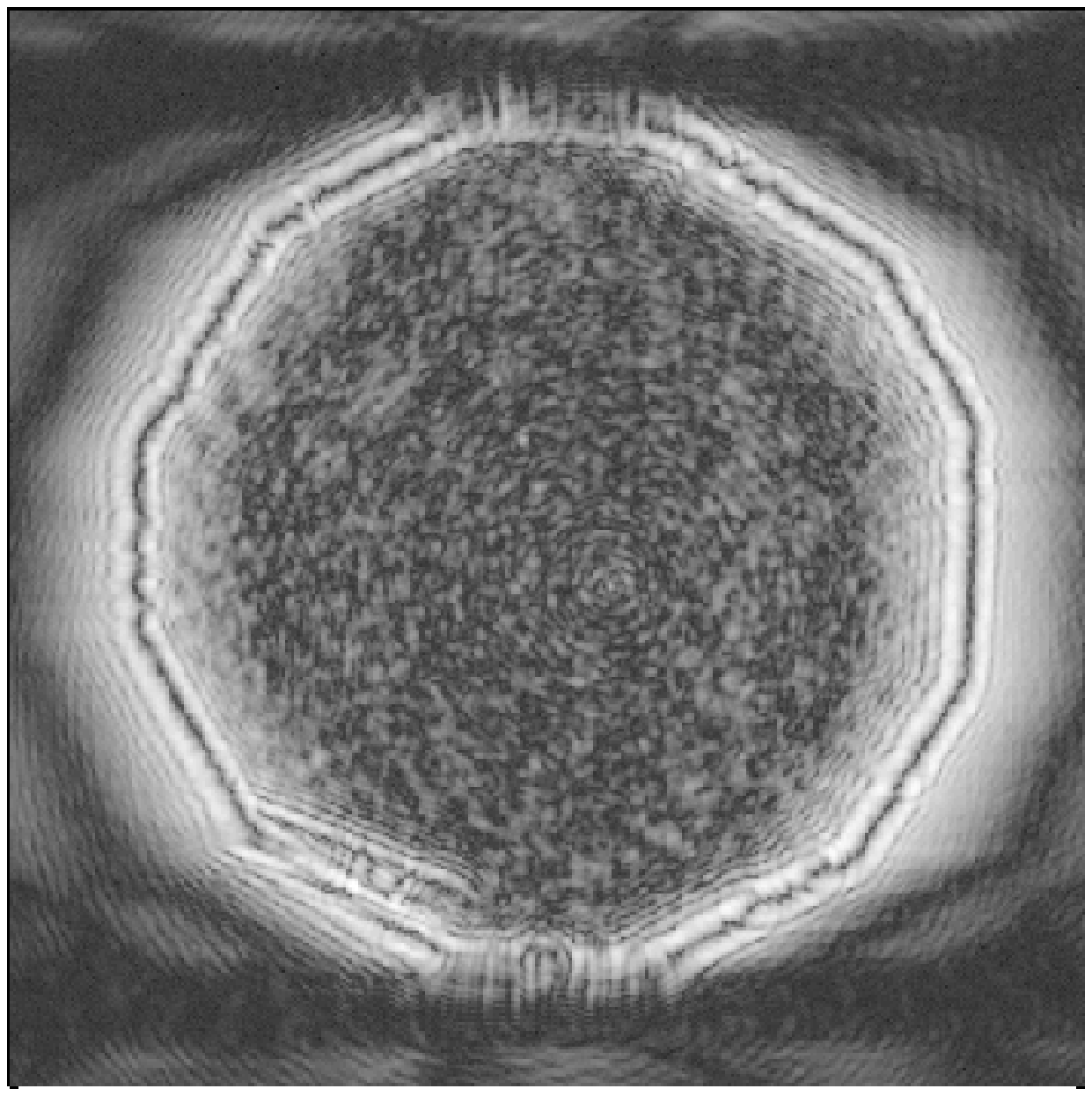}
\hfill
\includegraphics[height=2.1in]{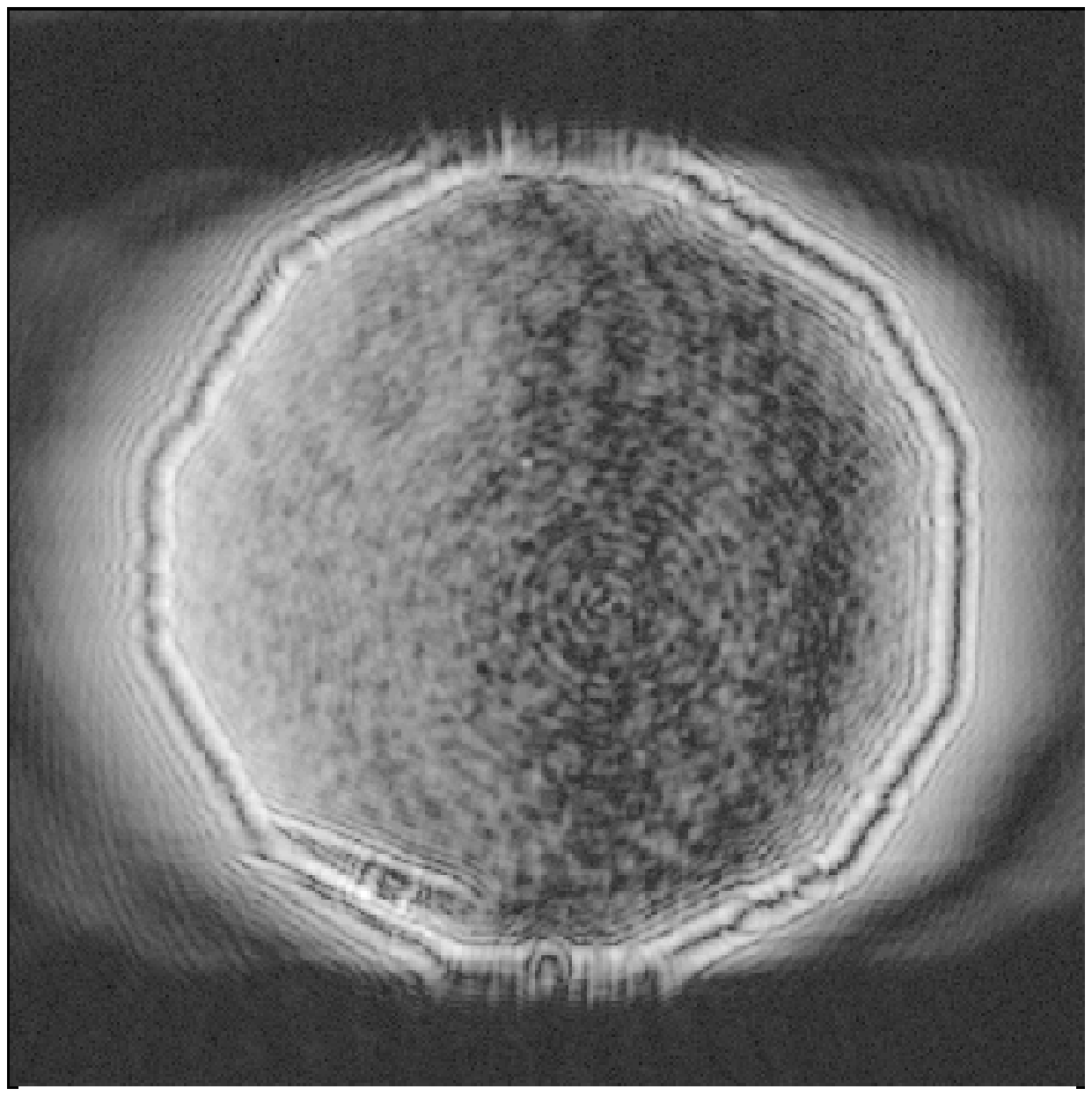}
\hfill
\raisebox{0.02in}[0pt]{\includegraphics[height=2.05in]{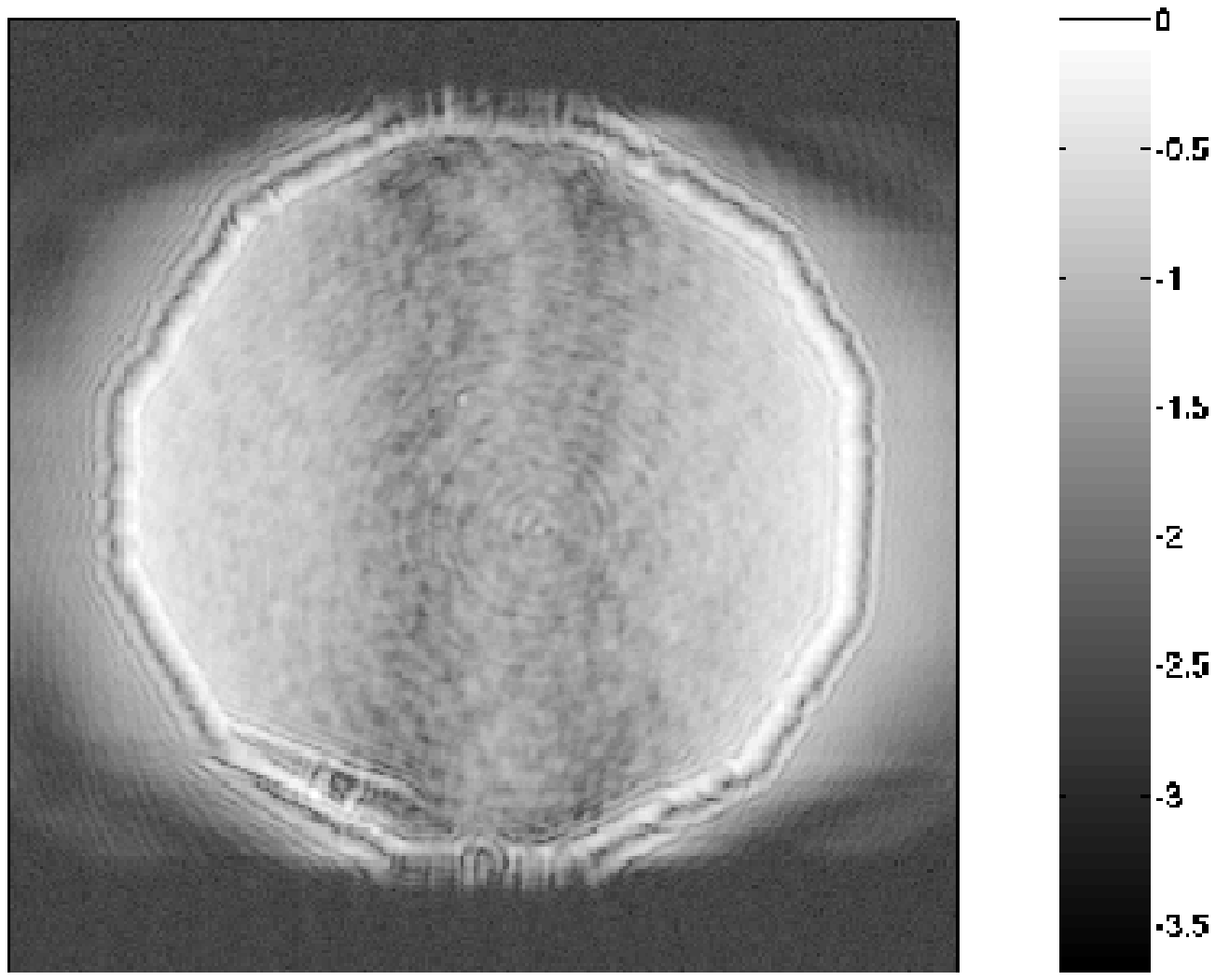}}
} 
\caption{Characteristic Lyot plane images with optimum mask alignment (left), and $2 \times 10^{-1}$ 
waves RMS tilt aberration (middle), and $9\times 10^{-2}$ waves RMS focus aberration (right) 
using the $m=1$, $l=2$ mask. Intensities are normalized to the peak intensity of the focus image
and are on a logarithmic scale. Shrinking the Lyot stop can improve contrast when certain aberrations are 
present, as is the case here. These images can be compared to the analytic predictions in Fig. 3 
of Sivaramakrishnan et al. 2005. Evidently, small amplitude tilt and focus phase aberrations 
produce a uniform leakage of light into the Lyot pupil interior. Although the aberrations presented here
are rather large, aspects of both images appear to reflect this phenomena. More complicated processes 
such as cross-talk between induced and inherent aberrations as well as frequency-folding from mask construction 
errors also contribute to the Lyot pupil field, and can create an intensity gradient.}
\label{fig:lyot_plane}
\end{figure}

For comparison, the SG05 theoretical tilt and focus curves are overplotted in 
Fig.~\ref{fig:aberr} (the triangular data points); these represent the steepest 
possible slopes that can be achieved in practice, since the model accounts only 
for diffraction. The general location of our experimental data are in good agreement 
with theory, and, to first order, simply adding a constant level of scattered
light to the SG05 diffracted light curves recovers the eighth-order
masks' experimental contrast to within the uncertainty of the
measurements. The fourth-order mask however makes exception, by
providing better contrast in practice than expected from theory. This
apparent discrepancy is resolved by considering a subtle difference
between the two studies: the Lyot stop size;\footnote{Other differences 
between this study and SG05 are: (1) we used a circular entrance aperture, 
instead of an elliptical entrance aperture, and (2) the contrast was measured 
at $3 \, \lambda / D$, instead of $4 \, \lambda / D$.} the SG05 simulations 
maximize off-axis throughput by choosing the largest possible Lyot stop
shape, whereas we have undersized the Lyot stop in this experiment
by $\sim 25\%$ for each mask. In the presence of aberrations, the contrast of a 
band-limited mask depends upon the size of the Lyot stop. If the light diffracted 
in the Lyot plane due to aberrations is non-uniform and less intense near the optical 
axis, decreasing the size of the Lyot stop can improve contrast, at a cost
of throughput and resolution (see Sivaramakrishnan et al. 2005). We find that the 
introduction of tilt and focus aberrations produces patterns in the Lyot plane that fit 
this description (Fig.~\ref{fig:lyot_plane}), and that the undersizing of the Lyot stop 
is responsible for an enhanced resistance to aberrations. Furthermore, this effect is 
more pronounced with the fourth-order mask than the eighth-order mask, since low-order 
aberrations can be partially or completely filtered in the image plane before reaching 
the Lyot pupil. 

It is a good assumption that the fourth- and eighth-order mask phase 
aberration contrast curves (Fig.~\ref{fig:aberr}) do not intersect at 
more than one point for a given aberration: the slope of a mask's 
dependence on {\it any} low-order Zernike mode is always at least 
as steep for eighth-order masks as it is for fourth-order masks (SG05). Since 
we see the intersection in both graphs in Fig.~\ref{fig:aberr}, we conclude 
that the eighth-order masks can achieve better contrast than the fourth-order
mask when small levels of tilt and focus aberrations are present,
even though this regime was not available for direct measurement in
our experiment. At intermediate aberration levels, we do indeed see
an improvement in contrast with the eighth-order masks. This is more
evident with tilt, because the intersection between experimental
curves occurs well above the scattered light floor.

Reducing the amount of scattered light in the coronagraph by
approximately two orders of magnitude will enable a more reliable 
extrapolation of data to smaller aberrations with future work; 
however, polishing the optics with such precision is not feasible. 
The achromat doublets in this experiment are the same that were used 
in Debes et al. 2004, with rms surface roughnesses of $\lesssim 1$ nm 
on spatial frequency scales that correspond to the search area. A better 
technique to compensate for the scattered light present in the system at 
this level would be to implement a deformable mirror (e.g. Trauger et al.
2004).

\section{SUMMARY}
We have built and tested three eighth-order notch filter masks and
one fourth-order notch filter mask - each with the same IWA - to
make a comparitive study of low-spatial-frequency optical aberration
sensitivities within the context of Lyot coronagraphy, using
monochromatic light. We find that the eighth-order masks are less
susceptible to the low-order aberrations of tilt and focus than the
fourth-order mask: they provide the same contrast as the
fourth-order mask in a system with either $\sim 5$ times as large a
pointing error or $\sim 4$ times as large an RMS focus wavefront
error. Additionally, the eighth-order masks show a stronger dependence 
to both tilt and focus at large aberration levels (i.e. a steeper slope), 
as predicted by theory. There was excellent agreement with our results 
and the SG05 numerical model, once the differences in each study were 
accounted for. We were unable to extrapolate our data to calculate the exact
aberration levels necessary to achieve $\lesssim 10^{-10}$ contrast
at the IWA because of the amount of scattered light in the system; doing 
so would require implementing a deformable mirror to reduce the scattered 
light levels by approximately two orders of magnitude. Transmission of light 
directly through the Chromium occulting layer accounted for $\sim 10\%$ of 
the noise floor at the IWA, but significantly less in the extended search 
area.

%area where there is significantly less off-axis attenuation in the image plane.

With ``perfect'' alignment, we find that all of the masks generate
contrast levels of $\sim 2 \times 10^{-6}$ at $3 \, \lambda / D$ and
$\sim 6 \times 10^{-7}$ at $10 \, \lambda / D$. In essence, the
on-axis (`stellar') flux was reduced by $7$ orders of magnitude at
the expense of attenuating off-axis light by a factor of $4-10$.
Since our system is ``diffraction limited'' (i.e. $\gtrsim 80\%$ Strehl ratio), 
we conclude that the $3 \sigma$ detection thresholds shown in 
Fig.~\ref{fig:contrast} represent the approximate deepest contrast that 
is achievable from the ground with current AO, using this type of coronagraph.

The Lyot stop throughput penalty in switching from a fourth-order
mask to an eighth-order mask was greatly exaggerated in this study,
because of nanofabrication limitations. We were able to achieve
$\sim 31\%$ Lyot stop throughput with the fourth-order mask and
$\sim 10\% \,-\, 16\%$ with the eighth-order masks. With a material
that is more opaque than Chrome at visible wavelengths, such as
Aluminum, eighth-order masks will be able to reach their full
potential in future experiments.

These results support the recent theoretical studies of Kuchner,
Crepp, \& Ge 2005 (KCG05) and Shaklan \& Green 2005 (SG05)
suggesting that eighth-order image masks can meet the demands of a
space mission designed to image extrasolar terrestrial planets, such
as the TPF-C, by providing the Lyot coronagraph with a large dynamic
range, high off-axis throughput, a large search area, and resistance 
to low-spatial-frequency wavefront aberrations.

% was more clear with tilt that the 8th-order did better; this
% is important for when people ask to overplot the other curve

% Not bad to be able to reduce the stellar flux by 7 orders of magnitude
% at a cost of decreasing the planet light by a factor of 4-10.

% we never tested the dependence of the contrast on the f/#

% Talk about the scratch/dig and compare to Lay ... also maybe mention
% fabrication errors? These effects are $\leq 10^{-8}$ effects though. Maybe
% we can discuss the problem of deciding whether graded or binary masks
% are better?

We thank Xiaoke Wan and Bo Zhao for their assistance in the lab,
Lakshminarayan Hariharan for building the optimized Lyot stops, and
Stuart Shaklan and John Debes for their helpful comments. We acknowledge
support from the NSF with grants NSF AST-0451407 and AST-0451408, NASA
with grants NNG05G321G and NNG05GR41G, the JPL TPF program, the UCF-UF SRI
program, the University of Florida and Penn State. This research was also funded
in part by a Grant-In-Aid of Research from the National Academy of Sciences,
Administered by Sigma Xi, The Scientific Research Society.


\begin{thebibliography}{}
\begin{small}

\bibitem[Aime, C. 2005]{aime05}Aime, C. 2005, A\&A, 434, 785
\bibitem[Debes et al. 2004]{debes04}Debes, J.~H., Ge, J., Kuchner, M.~J. Rogosky, M. 2004, \apj, 608, 1095
\bibitem[Ford et al.(2004)]{ford04} Ford, V.~G., Lisman, P.~D., Shaklan, S.~B., Trauger, J.~T., Ho, T., Hoppe, D., \& Lowman, A.~E.\
2004, \procspie, 5487, 1274
\bibitem[Kuchner 2004]{kuch04}Kuchner, M. J. astro-ph/0401256
\bibitem[Kuchner, Crepp, \& Ge 2005]{kcg05}Kuchner, M.~J., Crepp, J.~R., Ge, J. 2005, \apj, 628, 466 (KCG05)
\bibitem[Kuchner \& Spergel 2003]{kuch03}Kuchner, M.~J. \& Spergel, D.~N. 2003, \apj, 594, 617
\bibitem[Kuchner \& Traub 2002]{kuch02}Kuchner, M.~J. \& Traub, W.~A. 2002, \apj, 570, 900
\bibitem[Lay et al. 2005]{lay05}Lay, O.~P., Green, J.~J., Hoppe, D.~L., Shaklan, S.~B. 2005, Proc. SPIE, 5905, 148
\bibitem[Lloyd et al. 2001]{lloyd01}Lloyd, J. P., Graham, J. R., Kalas, P., Oppenheimer, B. R., Sivaramakrishnan, A., Makidon, R., Macintosh, B. A., Max, C., 
Baudoz, P., Kuhn, J., \& Potter, D. 2001, Proc. SPIE, 4490, 290
\bibitem[Lloyd \& Sivaramakrishnan]{lloyd05}Lloyd, J. P. \& Sivaramakrishnan, A. 2005, \apj, 621, 1153
\bibitem[Lopez, Schneider, \& Danchi 2005]{lsd05}Lopez, B., Schneider, J., Danchi, W.~C. 2005, \apj, 627, 974
\bibitem[Lyot 1939]{lyot}Lyot, B. 1939, MNRAS, 99, 580
\bibitem[Rouan 2004]{rouan04} Rouan, D. 2004, EAS Publications Series, 12, 21
\bibitem[Semaltianos 2001]{sem01}Semaltianos, N.~G. 2001, Applied Surface Science, 183, 223
\bibitem[Shaklan \& Green 2005]{sg05}Shaklan, S. B., \& Green, J. J. 2005, \apj, 628, 474 (SG05)
\bibitem[Sivaramakrishnan et al. 2005]{siv05}Sivaramakrishnan, A., Soummer, R., Sivaramakrishnan, A.~V., Lloyd, J.~P., Oppenheimer, B.~R.,
\& Makidon, R.~B. 2005, \apj, 634, 1416
\bibitem[Sivaramakrishnan et al. 2001]{siv01}Sivaramakrishnan, A., Koresko, C.~D., Makidon, R.~B., Berkefeld, T., Kuchner, M.~J. 2001, \apj, 552, 397
\bibitem[Trauger et al. 2004]{trau04}Trauger, J. et al. 2004, Proc. SPIE, 5487, 1330

\end{small}
\end{thebibliography}
\end{document}